\documentclass[3p]{elsarticle}

\usepackage{amssymb}
\usepackage{stackengine}
\usepackage[utf8]{inputenc}
 \usepackage{mathtools}
\usepackage{amssymb}
\usepackage{tikz}
\usepackage{bbding}
\usepackage{pifont}
\def\abs{\@ifstar{\oldabs}{\oldabs*}}
\usepackage{rotating}

\usepackage{nomencl}
\usepackage{algorithm,algpseudocode}
\usepackage{tgcursor}
\usepackage{amsmath,amsfonts,mathtools}
\usepackage{cite}
\usepackage{ wasysym }
\usepackage{multirow}
\usepackage{breqn}
\usepackage{array}
\usepackage{colortbl}
\usepackage{mdframed}
\usepackage{longtable}

\usepackage {hyperref}
\hypersetup{
colorlinks=true, 
linkcolor=blue, 
citecolor=blue, 
filecolor=magenta, 
urlcolor=blue,
bookmarksdepth=4
}

\definecolor{shadecolor}{rgb}{1,0.8,0.3}
\definecolor{lightgray}{rgb}{0.9,0.9,0.9}
\definecolor{yellow}{rgb}{1,0.5,0}

\usepackage{hyperref}
\hypersetup{
    colorlinks = false,
    linkbordercolor = {1 0 0}
}

\usepackage{framed} 

\usepackage{multicol} 

\usepackage{nomencl} 

\makenomenclature

\renewcommand*\nompreamble{\begin{multicols}{2}}

\renewcommand*\nompostamble{\end{multicols}}

\usepackage{etoolbox}
\renewcommand\nomgroup[1]{%
  \item[\bfseries
  \ifstrequal{#1}{I}{\textit{Indices and sets}}{%
  \ifstrequal{#1}{A}{\textit{Abbreviations}}{%
  \ifstrequal{#1}{B}{\textit{Superscripts}}{%
  \ifstrequal{#1}{P}{\textit{Parameters}}{%
  \ifstrequal{#1}{V}{\textit{Variables}}{
  \ifstrequal{#1}{Z}{\textit{Functions}}}}}}}%
]}
 
\usepackage{accents}

\usepackage{blkarray}

\begin{document}
\begin{frontmatter}

\title{Lightweight Blockchain Framework for Location-aware Peer-to-Peer Energy Trading}

\author[1]{Mohsen Khorasany\corref{cor1}%
\fnref{fn1}}

\author[2]{Ali Dorri
\corref{cor2}}
\author[1]{Reza Razzaghi}
\author[2]{Raja Jurdak}
\cortext[cor1]{Corresponding author; Department of Electrical and Computer Systems Engineering, Monash University, Clayton VIC 3800, Australia, Email: mohsen.khorasany@monash.edu}
\cortext[cor2]{Contributing equally with the first author}
\address[1]{Department of Electrical and Computer Systems Engineering, Monash University, Clayton VIC 3800, Australia}
\address[2]{School of Computer Science, Queensland University of Technology, Brisbane City, QLD 4000, Australia}

\begin{abstract}
Peer-to-Peer (P2P) energy trading can facilitate integration of a large number of small-scale producers and consumers into energy markets. Decentralized management of these new market participants is challenging in terms of market settlement, participant reputation and consideration of grid constraints. This paper proposes a blockchain-enabled framework for P2P energy trading among producer and consumer agents in a smart grid. A fully decentralized market settlement mechanism is designed, which does not rely on a centralized entity to settle the market and encourages producers and consumers to negotiate on energy trading with their nearby agents truthfully. To this end, the electrical distance of agents is considered in the pricing mechanism to encourage agents to trade with their neighboring agents. In addition, a reputation factor is considered for each agent, reflecting its past performance in delivering the committed energy. Before starting the negotiation, agents select their trading partners based on their preferences over the reputation and proximity of the trading partners. An Anonymous Proof of Location (A-PoL) algorithm is proposed that allows agents to prove their location without revealing their real identity. The practicality of the proposed framework is illustrated through several case studies, and its security and privacy are analyzed in detail. 

\end{abstract}




\end{frontmatter}


\section{Introduction}
In response to climate change concerns, the decarbonization and digitization of the electricity sector have been accelerated in recent years. The path toward decarbonization is associated with the high penetration of Distributed Energy Resources (DERs), such as rooftop solar, home batteries, and Electric Vehicles (EVs) with the potential to support a reliable, affordable and lower-emissions electricity grid. Progressive deployment of DERs raises several opportunities and challenges in the electricity systems \citep{openenergy}. Technical challenges include the increase in bidirectional power flows, the raise in the voltage levels, and lack of visibility \citep{coster2010integration}. On the other side, efficient DER integration can provide substantial benefits for energy customers \citep{aemo}. DER owners can benefit from Feed-In-Tariffs (FiTs) programs by selling energy back to the utility grid for a fixed price \citep{ye2017analysis}. Alternatively, they can be coordinated and orchestrated for participation in different markets.

\begin{figure}[t!]
	\centering
	\includegraphics[scale=0.5]{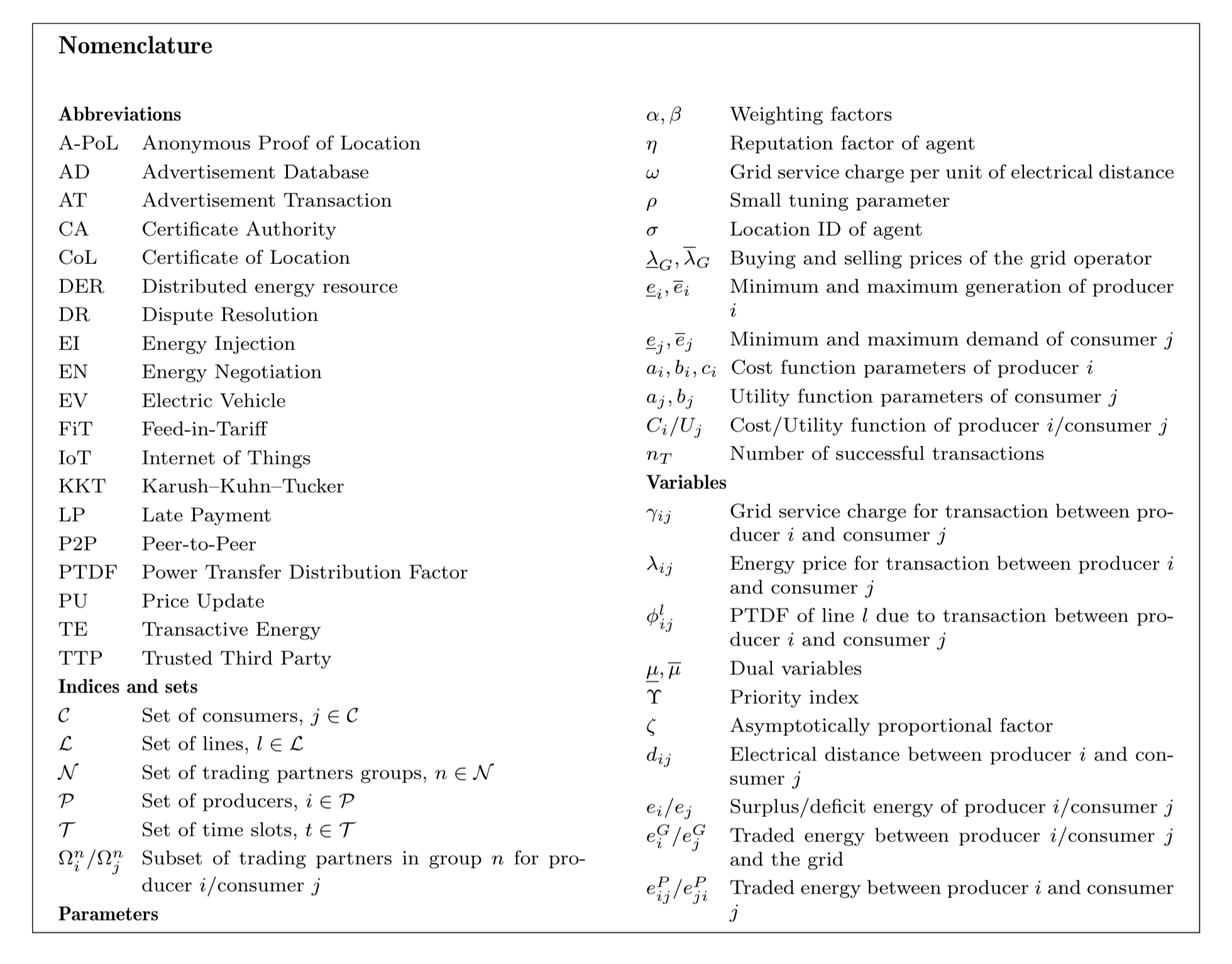}
\end{figure}

Given this context, Transactive Energy (TE) is emerging as a new approach for orchestrating and coordinating operation of DERs. TE is a market-based approach for energy management, which uses price signals to coordinate demand and supply across the network and among all users and entities \citep{council2015gridwise}. TE approach facilitates integration of DERs in the grid, while maintaining the system reliability \citep{liu2017transactive}. It provides a transformative solution to technological and socioeconomic challenges of the power grid modernization \citep{li2019blockchain}. Built upon the context of TE, a novel technique for the integration of small-scale producers and consumers to energy markets is Peer-to-Peer (P2P) trading, which allows bilateral energy transactions between users \citep{tushar2020peer}. P2P trading provides significant benefits for both end users and grid operators such as increasing welfare by preference satisfaction \citep{morstyn2018using}, lowering operational costs, and improving system reliability \citep{mengelkamp2018designing}. P2P trading offers the flexibility required to coordinate agents with diverse preferences. Recent advances in the Internet of Things (IoT) and digital technologies have paved the path toward grid digitization. Grid digitization provides a two-way communication network that allows DER owners and energy consumers to act as proactive agents in the energy markets and facilitates P2P market implementation. Through direct interaction of agents in a decentralized platform, small-scale producers are allowed to compete with large traditional suppliers, and consumers have the freedom to select their energy suppliers based on their preferences. 

In the P2P trading, it is expected that market participants can settle the bilateral transactions with least influence from a central authority \citep{giotitsas2015peer}. Hence, designing a decentralized platform for managing market participants with diverse preferences is a challenging task. Blockchain technology offers new opportunities for decentralized market designs due to its salient features which includes decentralization, trust, anonymity, and auditability. Blokchain enables energy producers and consumers to directly negotiate and trade energy without reliance on Trusted Third Party (TTP). It provides a platform to store and share data in a secure and verifiable manner, even when the identity and trustworthiness of market participants are unknown \citep{van2020integrated}. The participating nodes in the blockchain, that includes energy consumers, producers, prosumers, or grid operators, jointly form an overlay network where they can exchange information  \citep{wu2019comprehensive}. Given the properties of blockchain, and the need for a truly decentralized platform for P2P trading, designing blockchain-enabled frameworks for P2P trading is gaining momentum in recent years.

The blockchain framework for P2P trading should incorporate an appropriate market settlement approach to match trading peers and to settle the bilateral transactions among them. Compared to traditional markets, P2P market offers more flexibility and trading options to the agents, and hence, it needs a pricing mechanism that incentivizes both producers and consumers to participate in the market. There are several approaches that can be applied to P2P market clearing and peer matching, such as auction-based mechanism, game theory, and optimization-based methods \citep{li2018location, paudel2018peer, khorasany2019decentralised}. In the auction-based mechanism, agents express their interest in energy trading by submitting their offers, and the energy allocation and price would be determined based on the market clearing rules \citep{li2018location}. The game theory approaches aim to provide a stable solution that is beneficial for all parties \citep{paudel2018peer}. In the optimization-based methods, the market settlement is formulated as an optimization problem, which can be decomposed to local subproblems solvable by each agent \citep{khorasany2019decentralised}. The optimization-based methods can be implemented in a decentralized manner without any need for third party interventions, which allows agents to optimize their welfare by participating in the P2P market. Hence, these methods are well-suited for blockchain-enabled P2P markets. However, the computation and communication overheads of these methods are of concern, as they require agents to iteratively negotiate to reach agreements on their actions. Therefore, reducing these overheads is a key requirement

This paper designs a blockchain-enabled P2P market that provides a secure and transparent environment for the energy trading of producer and consumer agents. In the proposed approach, agents \textit{Advertise} their surplus/deficit energy during each market interval. We use an Advertisement Database (AD) that is centrally managed by the grid operator to skip storing advertisement in public blockchain, which in turn, reduces the blockchain memory footprint. A decentralized optimization algorithm is employed for the \textit{Negotiation} that allows agents to iteratively optimize their welfare in the negotiation process. 
In order to reduce the computation and communication overheads, a \textit{Prioritization} step is considered in the market settlement process that enables agents to prioritize their trading partners based on their proximity and reputation factor. Network constraints should be incorporated in the P2P trading to ensure that energy transactions respect electric grid constraints. Instead of enforcing network constraints directly to the proposed framework, we define a grid service charge for each transaction. To incentivize agents to trade energy with their neighboring agents and reducing network loading, this charge is calculated based on the electrical distance between producer and consumer. To this end, we propose an Anonymous Proof of Location (A-PoL) algorithm that enables agents to prove their location without revealing their real identity. Once the energy consumer and producer agree on the conditions of the trade, they start trading the energy. To reduce the reliance on TTP yet ensure the security of the trade, we adopt the concept of atomic meta-transactions~\citep{dorri2019spb} where two transactions are considered valid only if they are generated within a particular period of time. The contributions of this paper are summarized as follows:
\begin{itemize}
    \item [-] a decentralized P2P trading framework that does not require access to private information of agents in any stage of the market settlement;
    \item[-] a novel \textit{prioritization} step to allow agents to select their trading partners based on their location and reputation in order to reduce the system overheads;
    \item[-] a new A-PoL algorithm, which uses a Certificate of Location (CoL) issued by smart meters to approve the location of agents without revealing their real identity.
\end{itemize}

The rest of this paper is organized as follows. Section \ref{sec:related} explains some related work. Section \ref{sec: market structure} outlines the structure of the market, including agents modeling, market objective, and decentralized optimization of the market objective. Section \ref{sec:energy trading} explains the market settlement process and its different steps. Case studies and numerical results are reported in Section \ref{sec: case study}. A detailed analysis of the security and privacy of the proposed framework is presented in Section \ref{sec:security}. Finally, concluding remarks are given in Section \ref{sec:conclusion}.

\section{Prior Art}\label{sec:related}
In recent years blockchain applications in energy systems, such as demand response, microgrid energy management, EV integration, and energy trading, has attracted tremendous attention due to its salient features including decentralization, transparency, trust, and immutability \citep{andoni2019blockchain,musleh2019blockchain}. 
In \citep{van2020cooperative}, a decentralized framework for cooperative demand response is presented, in which smart contracts are employed to allow participants to collaboratively decide on their planning profiles. Su \textit{et. al} \citep{su2018secure} employed blockchain to implement secure charging services for EVs with the execution of smart contracts. In \citep{noor2018energy}, blockchain technology is utilized to implement a demand side energy management method for the efficient operation of microgrids. Hassan \textit{et. al} \citep{hassan2019deal} developed a blockchain-based approach for microgrid energy auction, in which to reduce computational complexity, at every node consortium blockchain is used that authorizes only selected nodes to add a new block in the blockchain.

The application of blockchain for P2P energy trading has been investigated in several studies. The Brooklyn microgrid is a prototype of a blockchain-enabled P2P market, in which a blockchain framework is employed to implement a decentralized microgrid energy market \citep{mengelkamp2018designing}. A unified blockchain framework for various scenarios of energy trading in an industrial IoT is proposed in \citep{li2017consortium}, in which a credit-based payment scheme is employed to overcome the transaction limitation caused by transaction confirmation delays. An operation framework for P2P energy trading at the distribution level is presented in \citep{wang2019energy}, where the system operator clears the market using a crowdsourced energy system model. Dang \textit{et. al} \citep{dang2019demand} proposed a blockchain-based P2P market for optimal load management of big industrial users, in which users can organize their own P2P market to save their electricity costs. In \citep{luo2018distributed}, a two-layer system for distributed electricity trading among prosumers is proposed, in which in the first layer prosumers can form coalitions and negotiate energy trading, and in the second layer blockchain is employed for settlement of transactions formed in the first layer. A methodology for the co-simulation of power distribution networks and local P2P energy trading platforms is presented in  \citep{hayes2020co}, in which a blockchain-based distributed double auction trade mechanism is employed to model the trading platform.

The existing blockchain-based frameworks for energy trading suffer from: \textit{(i) negotiation overheads:} the decentralized optimization methods rely on iterative negotiation between involved agents. In a market with large number of participants, this iterative process increases the communication and computation overheads, and consequently the negotiation time. \textit{(ii) Reliance on a TTP:} most of the existing studies rely on a TTP to oversee the trade and ensure that both sides of the trade commit to their obligations. However, this potentially may lead to centralization and privacy concerns as  TTP can build virtual profile about the users. \textit{(iii) Blockchain overheads:} in conventional blockchain-based methods in smart grids, all communications are stored in the blockchain, which in turn increases the blockchain memory footprint and thus reduces scalability. In this regard, this paper proposes a blockchain-enabled P2P market that alleviates the above-mentioned limitations.

\section{The Market Structure}\label{sec: market structure}
\begin{figure}[tb!]
    \centering
    \includegraphics[scale=0.8]{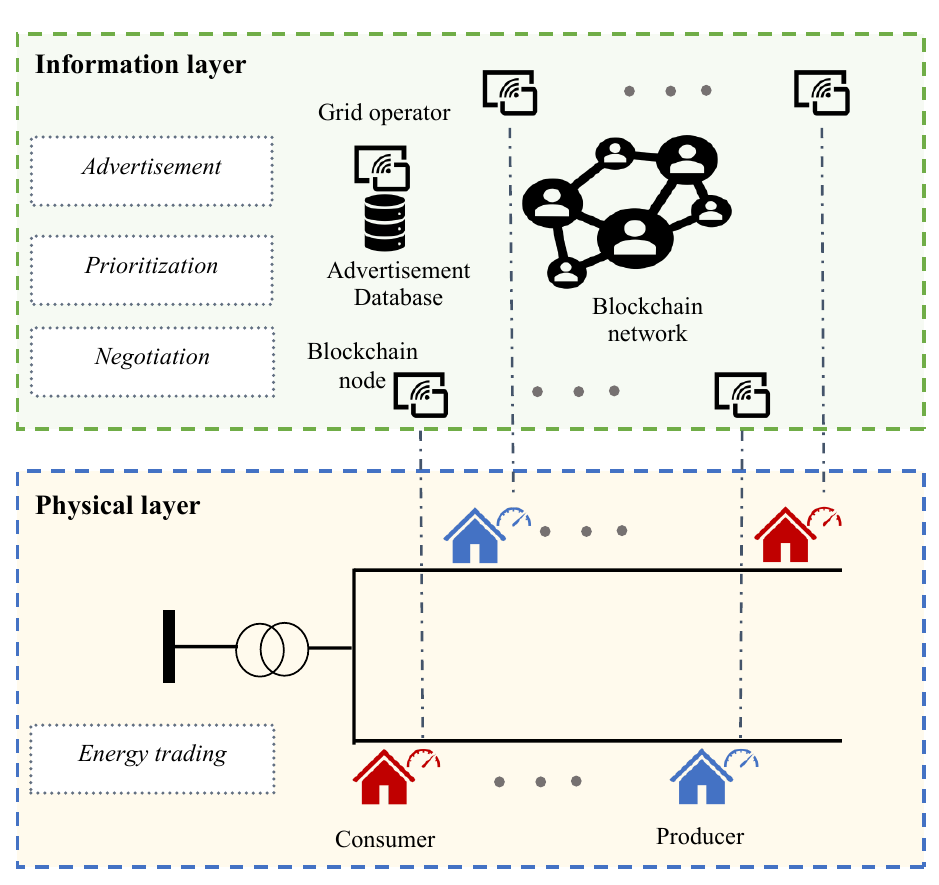}
    \caption{An overview of the proposed blockchain-enabled P2P energy trading.}
    \label{fig:layers}
\end{figure}
In this section, we outline the details of the market structure. As shown in Fig. \ref{fig:layers}, the proposed architecture consists of two layers: i) physical layer that is the underlying physical network to transfer electricity from producers to consumers.  The minimum requirement for a successful transaction between a producer and a consumer is the existence of a path to transfer power between them, and ii) information layer where the participating nodes, that includes energy producer, consumer, and the grid operator, connect through a public blockchain to share information. The information layer provides a secure platform for the participating nodes to advertise their energy, negotiate on their offers, and decide on their actions in the market. The market mechanism is implemented in this layer that enables the agents to trade energy and settle energy transactions.  \par 


\subsection{Agents modeling}\label{sub:sec:agent-modeling}
We consider an energy network with a set of agents consisting of a set of producer agents with index \(i \in \mathcal{P}=\{1, ..., P\}\) and a set of consumer agents with index \( j \in \mathcal{C}=\{1, ..., C\}\) connected to a physical layer manged by a grid operator. Producers have energy producing capability and can use their own generated energy for their demand. In case there is more generation than demand, producers can sell surplus energy to consumers or to the grid. Consumers, on the other side, can buy energy from producers or the grid. Producers and consumers can negotiate for energy trading in a forward market for any time interval \(t \in \mathcal{T} =\{1, ..., T\}\) with equal duration (e.g., one hour). Agents are equipped with smart meters, which determines the energy surplus/deficit for trade in each time slot. The smart meters are connected to the information layer and thus, can exchange information in blockchain. It is assumed that the smart meters are tamper-resistant. 
We assume that each agent is equipped with an energy management system that
can quantitatively estimate the energy surplus or deficit that needs to be traded in each market interval. Each agent can trade with the grid or with other agents in the network. The total energy surplus and deficit of producers and consumers is represented as
\begin{equation}\label{tot energ pro}
    e_i=e_i^G +\sum_{j \in \mathcal{C}}{e_{ij}^P}
\end{equation}
\begin{equation}\label{tot energ cons}
    e_j= e_j^G+\sum_{i \in \mathcal{P}}{e_{ji}^P}
\end{equation}
where, \(e_i^G\) and \(e_j^G\) are the traded energy between producer $i$/consumer $j$ and the grid respectively, \(e_{ij}^P\) is the sold energy by producer $i$ to consumer $j$, and \(e_{ji}^P\) is the purchased energy by consumer $j$ from producer $i$. Each agent in the P2P market aims to maximize its welfare. The welfare incorporates the utility/cost of energy consumption/generation, cost/revenue of trade with other agents or the grid, and the cost of using grid for the P2P trading.  The welfare function of producers and consumers can be modeled by (\ref{pro welf}) and (\ref{con welf}), respectively
\begin{equation}\label{pro welf}
    W_i(e_i, \lambda_{ij},\gamma_{ij})= \underline{\lambda}^G e_i^G +\sum_{j \in \mathcal{C}}{e_{ij}^P (\lambda_{ij}-\gamma_{ij})}-C_i(e_i) 
\end{equation}
\begin{equation}\label{con welf}
    W_j(e_j, \lambda_{ij},\gamma_{ij})= U_j(e_j)- \overline{\lambda}^G e_j^G -\sum_{i \in \mathcal{P}}{e_{ij}^P (\lambda_{ij}+\gamma_{ij})}
\end{equation}
where \(\underline{\lambda}^G\) denotes FiT representing the price for selling energy to the grid; \(\lambda_{ij}\) is energy price in transaction between producer $i$ and consumer $j$; \(\gamma_{ij}\) is grid service charge for using grid infrastructure for this trade; \(\overline{\lambda}^G\) denotes the price for selling from the grid which is usually a fixed value over the time (e.g. time of use tariff). The grid selling and buying prices limit energy price in the P2P market, i.e. for any bilateral trade
\begin{equation}\label{price lim}
    \underline{\lambda}^G \leq \lambda_{ij} \leq \overline{\lambda}^G.
\end{equation}

The cost function of the producer represents the cost of generating energy \(e_i\) and can be modeled as \citep{grainger2003power}
\begin{equation}\label{cost-func-producer}
    C_i(e_i)=a_i e_i^2 +b_i e_i +c_i 
\end{equation}
where \(a_i, b_i\) and $c_i$ are positive constants, which can be adjusted by the producer to reflect the generation cost. Since producers usually have renewable generation resources with zero marginal costs, the cost function can represent the cost associated with battery degradation cost, if the producer needs to discharge its battery to sell the energy. On the other side, the utility function of a consumer represents its satisfaction level by consuming energy $e_j$ and can be modeled as \citep{samadi2010optimal}
\begin{equation}\label{cost-function-consumer}
    U_j(e_j)=
    \begin{cases}
     -a_j e_j^2 +b_j e_j \;\;\;\; :0 \leq e_j \leq \frac{b_j}{2 a_j}\\
     \frac{b_j^2}{4 a_j}  \;\;\;\;\;\;\;\;\;\;\;\;\;\;\;\;\;\;\ :e_j \geq \frac{b_j}{2 a_j}
    \end{cases}
\end{equation}
where \(a_j\), and \(b_j\) are unique positive constants for each consumer. These parameters reflect the valuation of the energy by the consumer and denotes the price that consumer is willing to pay for the energy. A fundamental challenge in implementing P2P markets is how to deal with network constraints, without having a central dispatch mechanism. In this work, instead of enforcing network constraints directly, we use a grid service charge for each trade \citep{baroche2019exogenous}. This fee represents the price that agents need to pay for using the grid infrastructure for each trade. To incite agents to trade with their closest electrical partner and to reduce network loading, this charge is calculated based on the electrical distance between producer and consumer as in
\begin{equation}\label{service charge}
    \gamma_{ij}=\omega d_{ij} 
\end{equation}
where \(\omega\) is the grid service charge per unit of electrical distance for each unit of energy, and \(d_{ij}\) is the electrical distance between producer $i$ and consumer $j$. This distance can be calculated based on power transfer distance, which aggregates the absolute value of Power Transfer Distribution Factor (PTDF) induced by a trade as in
\begin{equation} \label{eq:ptdf}
    d_{ij}=\sum_{l \in \mathcal{L}}{\phi^l_{ij}}.
\end{equation}
For any trade, \(\phi^l_{ij}\) indicates the fraction of transacted power from producer $i$ to consumer $j$ that flows over a line $l$, and can be calculated using the method presented in \citep{wood2013power}. 
\subsection{Market objective} \label{sub:sec:market-settlement}
The market objective for P2P trading is formulated as social welfare maximization, which maximizes the total welfare of players in the market subject to the constraints, and mathematically can be modeled as:

\begin{equation}\label{tot objective}
\begin{aligned}
& \underset{\textbf{\textit{$e_i$,$e_j$}}}{\text{max}}
  \sum_{i \in \mathcal{P}}{W_i} + \sum_{j \in \mathcal{C}}{W_j}\\
  & \text{s.t. constraints.}
\end{aligned}
\end{equation}
As stated in (\ref{price lim}), the prices in the P2P market should always be beneficial for both producers and consumers. Hence, it is reasonable to assume that all agents try to maximize their traded energy in the P2P market and to minimize trading with the grid by setting $e_i^G=e_j^G=0$. Therefore, (\ref{tot objective}) can be rewritten as:
\begin{subequations}\label{social welfare}
\begin{equation}
\begin{aligned}
& \underset{\text{\textbf{{$e_{ij}$,$e_{ji}$}}}}{\text{max}}
  \sum_{j \in \mathcal{C}}{U_j(e_{ji})} - \sum_{i \in \mathcal{P}}{C_i({e_{ij}})} - \sum_{j \in \mathcal{C}}{\sum_{i \in \mathcal{P}}}{(e_{ij}+e_{ji})\gamma_{ij}}\\
\end{aligned}
\end{equation}
\begin{equation}\label{producer flex}
  \text{s.t.} \;\;\; \underline{e_i} \leq \sum_{j \in \mathcal{C}}{e_{ij}^P} \leq \overline{e_i} \;\;\;\;\;\;\;\;\;\; :\underline{\mu_i}, \overline{\mu_i} 
\end{equation}
\begin{equation}\label{consumer flex}
      \underline{e_j} \leq \sum_{i \in \mathcal{P}}{e_{ji}^P} \leq \overline{e_j} \;\;\;\;\;\;\;\;\;\;\;\;\;\;\;\; :\underline{\mu_j}, \overline{\mu_j} 
\end{equation}
\begin{equation}\label{demand supply}
      e_{ij}^P=e_{ji}^P \;\;\;\;\;\;\;\;\;\;\;\;\;\;\;\;\;\;\;\;\;\;\;\;\;\; :\lambda_{ij}
\end{equation}
\end{subequations}
where (\ref{producer flex}) and (\ref{consumer flex}) represents the flexibility constraints of producer, and consumer, respectively. The constraint (\ref{demand supply}) represents the power balance in transaction between producer $i$ and consumer $j$. \(\underline{\mu_i}, \overline{\mu_i}, \underline{\mu_j}, \overline{\mu_j}, \lambda_{ij}\) are dual variables associated with these constraints.

\subsection{Decentralized optimization} \label{sub:sec:coordination}
In this paper, our aim is to solve (\ref{social welfare}) with only P2P communications to ensure data privacy of the agents. To do so, a decentralized optimization approach is employed to formulate the market settlement in P2P market \citep{khorasany2019decentralised}.
In this approach, dual variables are used to decouple constraints, and then, the problem is decomposed to local subproblems solvable by producers and consumers. The local subproblem is solved by deploying the sub-gradient projection method \citep{boyd2011distributed}. Each agent contributes to solving the global problem by updating its local decision variables. The set of decision variables for producers and consumers are \{\(\lambda_{ij}, e_{ij}^P, \underline{\mu_i}, \overline{\mu_i}\}\), and  \{\( e_{ji}^P, \underline{\mu_j}, \overline{\mu_j}\)\}, respectively. The market settlement approach is an iterative process, in which agents update their decision variables iteratively and exchange information without revealing their private information. The updates of the decision variables of agents are based on the Karush–Kuhn–Tucker (KKT) optimality conditions of the local problems, and can be developed using first order deviate of the relaxed problem as follows:\\
\(\forall i \in \mathcal{P}\)
\begin{subequations}\label{sell update dec}
\begin{equation}\label{sell price update}
    \lambda_{ij}^{k+1}=\left[\lambda_{ij}^{k}-\rho_{\lambda}^k(e_{ij}^{P,k}-e_{ji}^{P,k})\right]^{+}
\end{equation}
\begin{equation}\label{sell mu low update}
    \underline{\mu_i}^{k+1}=\left[\underline{\mu_i}^{k}+\rho_{\mu}^k(\underline{e_i}-e_i^k)\right]^{+}
\end{equation}
\begin{equation}\label{sell mu up update}
    \overline{\mu_i}^{k+1}=\left[\overline{\mu_i}^{k}+\rho_{\mu}^k(e_i^k-\overline{e_i})\right]^{+}
\end{equation}
\begin{equation}\label{sell energy update}
    e_{ij}^{{P,k+1}}= \left[e_{ij}^{{P,k}}+\zeta_{ij}^k(\tilde{e}_{ij}^{P,k+1}-e_i^k)\right]^{+}\\
\end{equation}
\begin{equation}\label{set point sell update}
  \tilde{e}_{ij}^{P,k+1}=\frac{\lambda_{ij}^{k+1}-\gamma_{ij}-\overline{\mu_i}^{k+1}+\underline{\mu_i}^{k+1}-b_i}{2 a_i}  
\end{equation}


\end{subequations}
\(\forall j \in \mathcal{C}\)
\begin{subequations}\label{buyer update dec}
\begin{equation}\label{buyer mu low update}
    \underline{\mu_j}^{k+1}=\left[\underline{\mu_j}^{k}+\rho_{\mu}^k(\underline{e_j}-e_j^k)\right]^{+}
\end{equation}
\begin{equation}\label{buyer mu up update}
    \overline{\mu_j}^{k+1}=\left[\overline{\mu_j}^{k}+\rho_{\mu}^k(e_j^k-\overline{e_j})\right]^{+}
\end{equation}
\begin{equation}\label{buyer power update}
    e_{ji}^{P,k+1}= \left[e_{ji}^{P,k}+\zeta_{ji}^k(\tilde{e}_{ji}^{P,k+1}-e_j^k)\right]^{+}
\end{equation}
\begin{equation}\label{buyer set point update}
    \tilde{e}_{ji}^{P,k+1}=\frac{b_j-\lambda_{ij}^{k+1}-\gamma_{ij}-\overline{\mu_j}^{k+1}+\underline{\mu_j}^{k+1}}{2 a_j}
\end{equation}
\end{subequations}
where $k$ is the iteration index, \(\tilde{e}_{ij}^{P}, \tilde{e}_{ji}^{P}\) are optimal power set points of producer and consumer at the price \(\lambda_{ij}\). \(\zeta_{ij}, \zeta_{ji}\) are asymptotically proportional factors, \(\rho\) is a small tuning parameter, and \([.]^+\) denotes max \{.,0\}. The information exchange between producers and consumers during the decentralized optimization process is explained in Section \ref{sec:neg}.

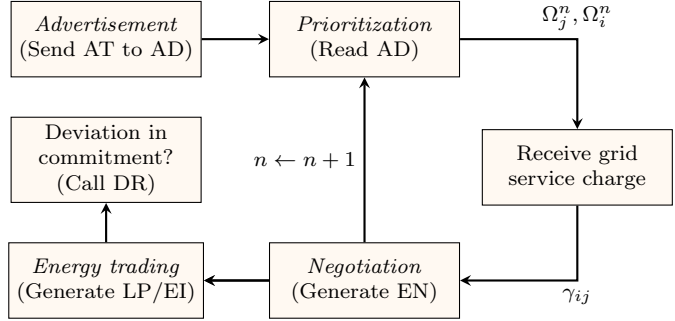
\begin{figure}[tb!]
\centering
\tikzstyle{startstop} = [rectangle, rounded corners, minimum width=2cm, minimum height=0.5cm,text centered, draw=black]

\tikzstyle{io} = [trapezium, trapezium left angle=70, trapezium right angle=110, minimum width=3cm, minimum height=0.5cm, text centered, draw=black]

\tikzstyle{process} = [rectangle, minimum width=2.5cm, minimum height=1cm, text centered, draw=black, fill=yellow!5]
\tikzstyle{decision} = [diamond, aspect=3, minimum width=0.1cm, minimum height=0.1cm, text centered, draw=black]
\tikzstyle{arrow} = [thick,->,>=stealth]

\begin{tikzpicture}[node distance= 1.4 cm, auto] 
\footnotesize{
\centering

\node (pro1) [process,align=center] {\textit{Advertisement}\\(Send AT to AD)};
\node (pro6) [process,align=center,yshift=-0.2cm,below of=pro1] {Deviation in\\ commitment?\\(Call DR)};
\node (pro2) [process, right of=pro1, xshift=2cm, align=center] {\textit{Prioritization}\\(Read AD)};
\node (pro3) [process, right of=pro2, xshift=1.4cm,yshift=-1.7cm,align=center] {Receive grid \\service charge};
\node (pro4) [process, below of=pro2, yshift=-1.8cm, align=center] {\textit{Negotiation}\\(Generate EN)};
\node (pro5) [process, below of=pro1, yshift=-1.8cm,align=center] {\textit{Energy trading}\\(Generate LP/EI)};

\draw [arrow] (pro1) -- node[anchor=south] {} (pro2);
\draw [arrow] (pro2) -|node[anchor=south] {$\Omega_j^n,\Omega_i^n$} (pro3);
\draw [arrow] (pro3) |-node[anchor=north] {$\gamma_{ij}$} (pro4);
\draw [arrow] (pro4) --node[anchor=south] {} (pro5);
\draw [arrow] (pro4) --node[anchor=north] {} (pro5);
\draw [arrow] (pro4) -- node[anchor=east] {$n \gets n+1$} (pro2);
\draw [arrow] (pro5) --node[anchor=north] {} (pro6);
}
\end{tikzpicture}
\caption{Market settlement algorithm.}\label{fig:full algorithm}
\end{figure}
\section{Market Settlement Process}\label{sec:energy trading}
In this section, we outline the details of the market settlement process for P2P trading. The proposed framework consists of four main phases namely;  \textit{(i) Advertisement:} to enable agents to advertise the energy that they want to trade, \textit{(ii) Prioritization:} to allow agent prioritize their trading partners based on their preferences, \textit{(iii) Negotiation:} in which agents negotiate on the energy quantity and price in each transaction, and \textit{(iv) Energy trading:} which is the step that energy transfer and payment would be taking place. These steps are summarized in Fig. \ref{fig:full algorithm} and discussed in detail in the rest of this section.

\subsection{Anonymous proof of location}
As shown in (\ref{service charge}) in the proposed framework the grid service charge is calculated based on the location of the participants denoted by $\sigma$ which requires the involved parties to reveal their location. However, this potentially enables the malicious nodes to track the activities of a particular user, which in turn compromises user privacy. Additionally, the distributed and anonymous nature of blockchain makes it challenging for the users to verify the location claimed by another node. To address these challenges, we propose an A-PoL algorithm that enables the users to prove their location while protecting their real identity, which in turn enhances the level of anonymity offered to the users. 

\begin{figure}[tb!]
    \centering
    \tikzstyle{startstop} = [rectangle, rounded corners, minimum width=2cm, minimum height=0.5cm,text centered, draw=black]

\tikzstyle{io} = [trapezium, trapezium left angle=70, trapezium right angle=110, minimum width=3cm, minimum height=0.5cm, text centered, draw=black]
\tikzstyle{Mytext} = [rectangle, minimum width=0.5cm, minimum height=0.2cm,text centered]
\tikzstyle{process} = [rectangle, minimum width=2.5cm, minimum height=1cm, text centered, draw=black, fill=yellow!5]
\tikzstyle{decision} = [diamond, aspect=3, minimum width=0.1cm, minimum height=0.1cm, text centered, draw=black]
\tikzstyle{arrow} = [thick,->,>=stealth]

\begin{tikzpicture}[node distance= 1.4 cm, auto] 
\footnotesize{
\centering

\node (pro1) [process,align=center] {Smart meter \textit{A}};
\node (pro2) [process, right of=pro1, xshift=2cm, yshift=1.7cm, align=center] {Certificate
\\Authority};
\node (pro3) [process, right of=pro2, xshift=2cm,yshift=-1.7cm,align=center] {Smart meter \textit{B}\\(Verifier)};
\node (pro4) [process, below of=pro2, yshift=-2cm, align=center] {Smart meter \textit{C}};
\node (text1) [Mytext,below of=pro2,align=center,yshift=-1.2cm,xshift=-0.6cm]{6. Verify};
\node (text2) [Mytext,right of=pro1,align=center,yshift=0.5cm,xshift=+1.cm]{2. Request CoL};
\node (text3) [Mytext,left of=pro3,align=center,yshift=-.5cm,xshift=-0.8cm]{4. Send CoL};
\node (text4) [Mytext,left of=pro2,align=center,yshift=.3cm,xshift=-1cm]{1. Establish PK};
\node (text5) [Mytext,right of=pro2,align=center,yshift=.25cm,xshift=+0.7cm]{3. Verify};
\node (text6) [Mytext,below of=pro1,align=center,yshift=-0.6cm,xshift=+1.1cm]{5. Share CoL};

\draw [arrow] (pro2) -| node[anchor=south] {} (pro1);
\draw [arrow] (pro2) -|node[anchor=south] {} (pro3);
\draw [arrow] (pro3) |- (pro2);
\draw [arrow] (pro1) |-node[anchor=north] {} (pro4);
\draw [arrow] (pro4) -- node[anchor=west] {} (pro2);

\draw [arrow] ([yshift=-3mm] pro3.west) -- node[anchor=south] {} ([yshift=-3mm] pro1.east);
\draw [arrow] ([yshift=3mm] pro1.east) -- node[anchor=north] {} ([yshift=3mm] pro3.west);
}
\end{tikzpicture}
    \caption{An overview of the proposed CoL.}
    \label{fig:CoL}
\end{figure}
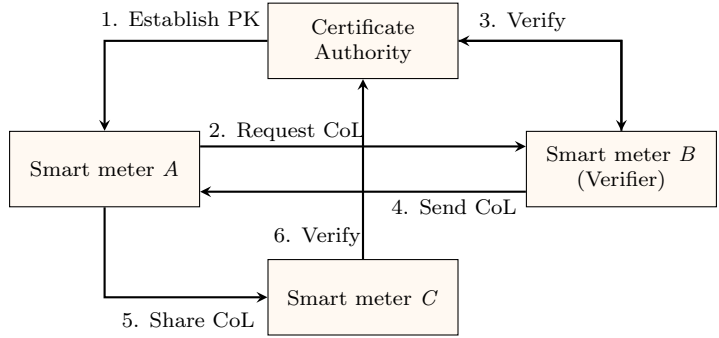

The proposed A-PoL algorithm involves a CoL that is issued by a smart meter in the network, as shown in  Fig. \ref{fig:CoL}. The energy companies maintain a local record of the accurate location of the installed smart meters. During the installation process, the energy company deploys a key pair in each smart meter (step 1 in Fig. \ref{fig:CoL}) and records $<PK,location>$ tuple in its local storage. The company serves as a certificate authority (CA) for PKs deployed in smart meters. Although the CA is a trusted authority, relying on the PK provided by the CA may  compromise the privacy of the users as the company can build a virtual profile of the user and their energy trading (by observing the proof of location transactions). To address this challenge, we propose CoL. \par 

CoL is a certificate received from a verifier that is an independent smart meter in the network. Assume smart meter \textit{A} is going to establish a CoL. Once deployed on site, \textit{A} explores the CA to find potential smart meters that can function as the verifier and selects one randomly. Assume smart meter \textit{B} is selected by \textit{A} to function as the verifier. Recall that we assume the smart meters are tamper resistant, and thus \textit{A} can send its request to any meter listed in CA.  \textit{A} sends a CoL request transaction to \textit{B} that is structured as $<T\_ID, MTR, \sigma, PK, Sign>$, where \textit{T\_ID} is the unique identifier of the transaction which essentially is the hash of the transaction content. \par

\textit{A} populates the root hash of a Merkle tree constructed by recursively hashing a number of PKs in the \textit{MTR} field. The PKs in the leaves of the tree are later employed by \textit{A} to prove ownership on the CoL, which is discussed in greater detail later in this section. The number of PKs may vary based on the application. $\sigma$ is the noisy location of the smart meter that can be the location at a lower resolution, e.g., the street in which the meter is installed. This  protects the privacy of the smart meter against deanonymization, as studied later in Section \ref{sec:security}. \textit{PK} is the PK of \textit{A} allocated by the CA, and \textit{Sign} is the corresponding signature, which proves that \textit{A} owns the private key corresponding to the PK. \par 

When the verifier, i.e. \textit{B} receives the CoL request, it verifies that the transaction is generated by a genuine smart meter that is done by requesting the CA (step 3). To protect the privacy of the user, the CA does not reveal the actual location of the smart meter to \textit{B} instead, only confirms if the PK is registered and genuine smart meter. Once verified, the verifier creates a CoL that is \textit{sign(hash(MTR, $\sigma$))} and replies back to \textit{A} by sending the reply transaction structured as  $<CoL, PK, Sign>$, where CoL is as outlined above, \textit{PK} is the PK of the verifier, i.e., \textit{B} registered by the CA, and \textit{Sign} is the corresponding signature of the verifier. \par

\textit{A} employs the received  CoL to anonymously prove its location to the nodes in the overlay. To do so, \textit{A} appends $CoL_f^{A} = (CoL_A, PK_{ver}, Sig_{ver}, MTR_A, \sigma_A, PK_A, MTL_A, Sign_A)$ where the first three fields are extracted from the response of the verifier \textit{B}. MTR and $\sigma_A$ are extracted from the CoL request sent to the verifier. \textit{$PK_A$} is a PK that was part of the leaves of the Merkle tree. $MTL_A$ is the leaves of the Merkle tree that are necessary to prove the existence of $PK_A$ in the MTR, and $Sign_A$ is the signature corresponding to PK. The last three fields ensure that only the owner of the certificate, who knows the PKs in the Merkle tree and the corresponding private key, can use the certificate. \par 

To verify the location of \textit{A}, the participating nodes in the blockchain, say \textit{C}, must first verify that $PK_A$ belongs to $MTR_R$ using $MTL_A$. Next, \textit{C} verifies if $Sign_A$ matches $PK_A$. The third step in verifying is to verify if $hash(MTR_A, \sigma_A) = CoL_A$. The final step is for \textit{C} to verify $PK_{ver}$ using CA. This ensures that a genuine smart meter has signed the CoL. If all the above steps successfully passed, the location of \textit{A} is verified. \par 

Having discussed the details of the proposed A-PoL algorithm, we study the process of P2P energy trading in the rest of this section. The algorithms implemented by producer and consumer agents are represented in Algorithm \ref{producer lag} and \ref{consumer alg}, respectively, and are discussed with greater details in the rest of this section. 

\subsection{Advertisement}\label{sub:sec:advertisement}
The main aim of this phase is for the agents to advertise their energy to the potential consumers/producers. In each market interval, agents participate in the forward market by submitting their offers and asks in the form of \textit{advertisement transaction (AT)} that is structured as follow $ AT = (T\_ID, price/amount, \eta_i, CoL_f^{i})$, where \textit{price/amount} can be either the price of the energy, i.e., $\lambda_i, \forall i\in \mathcal{P}$, if AT is generated by a producer or the amount of requested energy, i.e., $e_j, \forall j\in \mathcal{C}$,  if AT is generated by a consumer, and \(\eta\) is the reputation factor of agent.\par 

In conventional blockchain-based frameworks, the negotiation transactions are stored in the blockchain that provides high auditability. However, this increases the computational and storage resource consumption of the blockchain and limits the throughput, i.e., the total number of transactions that can be stored per second \citep{zhang2020privacy}. These can limit the blockchain scalability while smart grid comprises of a broad range of devices that may produce transactions frequently. To address this challenge, in the proposed framework, the negotiation transactions are stored in a public database managed by the grid operator that is referred to as \textit{Advertisement Database} (AD). The write access of AD is restricted only to the grid operator and other nodes in the overlay have read-only permission. The parties involved in an AT transaction, i.e., energy producers and consumers, may store the transactions locally to be employed as a reference in case of a dispute between parties. The final price and the amount of energy agreed by the participants are later defined during a negotiation which limits  future references to AT transaction. Thus, we store AT in AD which potentially reduces the packet overhead and blockchain memory footprint.

\begin{algorithm}[tb!]
\caption{Producer's Algorithm}\label{producer lag}
\footnotesize{
\begin{algorithmic}[1]
\State Submit offer AT \Comment{\textit{Advertisement}} 
\State Explore AD to find potential trading partners
\State Divide consumers in groups $\Omega_i^1, ..., \Omega_i^N$ using (\ref{prioritization sell}) \Comment{\textit{Prioritization}}
\State Set $n\gets 1$
\State Receive EN transactions from consumers \Comment{\textit{Negotiation}}
	\While{$|\lambda_{ij}^{P,k+1}-\lambda_{ij}^{P,k}| \geq \epsilon$}
		\For {$j \in \Omega_i$}
		\State Receive $\gamma_{ij}$ from grid operator
			\State Receive $e_{ji}^{k}$ from consumer
			\State Calculate $\lambda_{ij}^{k+1}$ using (\ref{sell price update})
			\State Update $\overline{\mu}_i^{k+1}$ and $\underline{\mu}_i^{k+1}$ using (\ref{sell mu low update}) and (\ref{sell mu up update})
			\State Calculate $e_{ij}^{P,k+1}$ using (\ref{sell energy update})
			\State Broadcast $\lambda_{ij}^{P,k+1}$ to consumers
		\EndFor
	\EndWhile
	\State Check if more energy is available
	\State Set $n\gets n+1$
	\State Repeat Negotiation with new consumers
	\State Receive LP from consumers \Comment{\textit{Energy trading}}
	\State Inject agreed energy
	\State Sign EI
\end{algorithmic}
}
\end{algorithm}

\subsection{Prioritization}
After \textit{advertisement} step, agents explore the available ATs in the AD for the negotiation. One-on-one negotiation can increase the delay associated with finalizing the trade as either side of the trade may have higher welfare in trading with another party in the network. Also, negotiation with every agent in the market increases the computation and communication overheads, which potentially leads to low scalability \citep{khorasany2019enhancing}. Thus, agents need to prioritize their trading partners based on their preferences and only negotiate with a given number of agents.

As stated in Section \ref{sub:sec:agent-modeling}, agents have to pay a grid service charge for each transaction as defined in (\ref{service charge}). This charge is directly associated with the distance between the producer \textit{i} and the consumer \textit{j}, and impacts the welfare of the agents. Accordingly, agents are incentivized to trade with trading partners located in close vicinity. This reduces the load on the transmission lines and thus reduces the cost in managing the physical layer. On the other hand, agents may prefer to negotiate with the trading partners with higher reputation factors, indicating their past performance in fulfilling their commitments. Thus, agents need to prioritize their trading partners based on their preferences to limit the partner search space. These preferences include reputation factor and distance. Agents may have varied preferences over reputation factor and distance of their trading partners. Hence, we define a priority index for each possible transaction between producers and consumers. This index for offer from consumer $j$ received by producer $i$, and offer from producer $i$ received by consumer $j$ are defined using (\ref{sell pri index}) and (\ref{buy pri index}), respectively.
\begin{equation}\label{sell pri index}
\Upsilon_{ij}=\alpha_i \rho_j +\beta_i (1-\cfrac{|\sigma_i -\sigma_j|}{D_{ij}})\\
\end{equation}
\begin{equation}\label{buy pri index}
\Upsilon_{ji}=\alpha_j \rho_i +\beta_j (1-\cfrac{|\sigma_i -\sigma_j|}{D_{ji}})\\
\end{equation}
where \(\alpha\) and \(\beta\) are the weights that agent places on the reputation factor and proximity of other agents, respectively such that \(\alpha+\beta=1\), \(D_{ij}=\underset{\sigma_j}{\max} |\sigma_i-\sigma_j|\), \(D_{ji}=\underset{\sigma_i}{\max} |\sigma_i-\sigma_j|\). The second term in (\ref{sell pri index}) and (\ref{buy pri index}) denotes the proximity of each trading partner to the agent, such that for the furthest trading partner of each agent this value is zero. After calculating priority index, each agent divides its trading partners to a set of $\mathcal{N}$ groups and then, starts to negotiate with the agents from the group with the highest priority and proceed to the group with the lowest priority. Each group of potential consumers/producers for producer $i$/consumer $j$ can be formed as
\begin{equation}\label{prioritization sell}
\Omega_i^n=\{j \in \mathcal{C}| (N-n)/N \leq \Upsilon_{ij} \leq (N-n+1)/N \} , \forall n \in \mathcal{N}
\end{equation}
\begin{equation}\label{prioritization buy}
\Omega_j^n=\{i \in \mathcal{P}| (N-n)/N \leq \Upsilon_{ji} \leq (N-n+1)/N \} , \forall n \in \mathcal{N}
\end{equation}
in which for consumer $j$, producers in group $\Omega_j^n$ have higher priority than producers in $\Omega_j^{n+1}$. Similarly, for producer $i$, consumers in group $\Omega_i^n$ have higher priority than consumers in $\Omega_i^{n+1}$.

\begin{algorithm}[tb!]
\caption{Consumer's Algorithm}\label{consumer alg}
\footnotesize{
\begin{algorithmic}[1]
\State Submit ask AT \Comment{\textit{Advertisement}}
\State Explore AD to find potential trading partners
\State Divide producers in groups $\Omega_j^1, ..., \Omega_j^N$ using (\ref{prioritization buy}) \Comment{\textit{Prioritization}}
\State Set $n\gets 1$
\State Send EN transactions to producers \Comment{\textit{Negotiation}}
	\While{$|e_j^{P,k+1}-e_j^{P,k}| \geq \epsilon$}
		\For {$i \in \Omega_j^n$}
		    \State Receive $\gamma_{ij}$ from grid operator
			\State Receive $\lambda_{ij}^{k}$ from producer
			\State Update $\overline{\mu}_j^{k+1}$ and $\underline{\mu}_j^{k+1}$ using (\ref{buyer mu low update}) and (\ref{buyer mu up update})
			\State Calculate $e_{ji}^{P,k+1}$ using (\ref{buyer power update})
			\State Broadcast $e_{ji}^{P,k+1}$ to producers
		\EndFor
	\EndWhile
	\State Check if more energy is needed
	\State Set $n\gets n+1$
	\State Repeat Negotiation with new producers
		\State Send LP to producers \Comment{\textit{Energy trading}}
	\State Sign EI
\end{algorithmic}
}
\end{algorithm}
\subsection{Negotiation}\label{sec:neg}
 
After \textit{prioritization}, each consumer starts negotiating with producer agents in $\Omega_j^1$. The first step is for consumer \textit{A}  to request the grid service charge from the grid operator. \textit{A} sends the list of agents in $\Omega_j^1$ to the grid operator who calculates grid service charges using (\ref{service charge}) and sends the response back to \textit{A}. Once \textit{A} receives the grid service charge, it starts the  negotiation with agents in $\Omega_j^1$ by sending an \textit{Energy Negotiation} (EN) transaction that is structured as $ <T\_ID, Amount, Price, PK^{D}, Sign^{D}, PK^{+}, Sign,  PK^{+},\\ Agreement_P, Agreement_C>$
where  \textit{Amount} identifies the total amount of energy demanded by \textit{A} and \textit{price} is the price in which the consumer is willing to pay.  \textit{PK\textsuperscript{D}} and \textit{ Sign\textsuperscript{D}} represent PK of the destination agent and its corresponding signature. The destination can be either the producer, when \textit{A} generates the transaction, or \textit{A}, when a producer is the generator of the transaction. This potentially enables routing the transaction to the destination using the routing algorithm proposed in \citep{dorri2019spb}. \textit{Sign } and \textit{PK}  represent the PK and signature of the transaction generator. Finally, \textit{$Agreement_P$} and \textit{$Agreement_C $} are flags that indicate whether the energy producer or consumer, respectively, agree with all situations in the trade. The generator of EN signs the hash of the transaction content which potentially ensures the integrity of the transaction content. After populating EN, \textit{A} sends the transaction to the energy producers in $\Omega_j^1$. \par

The energy producer may receive several EN transactions from consumers. The producer only responds to those who are in the set of $\Omega_i^1$. For each received EN from consumers in this set, the producer updates the price using (\ref{sell price update}).  When the response from the producer is received by the consumers, they update their energy using (\ref{buyer power update}) and again respond to the producer. This process continues till both sides of the trade agree on the trade conditions and thus set $Agreement_P$ and $ Agreement_C$ as '1'.  An EN transaction is considered as a valid transaction to be stored in the blockchain only when both energy producer and consumer sign the transaction and $Agreement_P$ and $ Agreement_C$ are set as '1'. This ensures that only the last transaction that contains the trade conditions is stored in the blockchain, which in turn increases the blockchain throughput and reduces the delay of the negotiation.

\subsection{Energy trading}
In this section, we discuss the energy trading process. Once agents agree on the trade condition, during the negotiation step, the consumer generates a \textit{Late Payment} (LP) transaction that pays the energy price to the producer. Conventional energy trading frameworks rely on a TTP to oversee the trade and ensure that both sides of the trade commit to their obligations, which in turn reduces the privacy of the users. To address this challenge, our framework relies on atomic meta-transactions. In the latter, two transactions are considered valid if and only if both are generated within a specified time frame. Any incompatible transaction is considered as invalid and thus is not stored in the blockchain \citep{dorri2019spb}. LP is an atomic meta-transactions thus the energy price is not transferred to the producer accoutn unless LP is coupled with another transaction that is discussed  in next paragraph. LP is structured as $<T\_ID, Price, Input, Output, EN\_Ref, Expiry\_Time, \\ Sign>$, where \textit{price} is price to be paid to the energy producer. \textit{Input} is the address of an unspent transaction that has enough balance to pay the transaction price, and \textit{output} is the address of the energy producer as in the last EN transaction. \textit{En\_Ref} is the address of the EN transaction that is stored in the public blockchain.  \textit{Expiry\_Time} represents the time period in which the second transaction corresponding to the current LP must be generated, otherwise, LP is discarded.  \textit{Sign}  is the signature of the transaction generator that must be corresponding to the PK used in EN transaction. The consumer then broadcasts LP transaction.\par 

The energy producer starts injecting energy to the grid when it receives the LP transaction. Once the total amount of agreed energy is injected to the grid,  the smart meter of the producer generates an \textit{Energy Injection} (EI) transaction which is a multisig transaction and is structured as $ <Amount, LP\_ID, PK\_P, Sign\_P,\\ PK\_C, Sign\_C>$, where \textit{Amount} is the amount of energy injected into the grid by the producer. \textit{LP\_ID} is \textit{T\_ID}  of the corresponding LP that is used for verification of the trade as outlined in the next paragraph. EI requires two out of two signatures to be considered as a valid transaction that are the energy producer signature, populated in \textit{PK\_P} and \textit{Sign\_P}, and energy consumer signature, populated in  \textit{PK\_C} and \textit{Sign\_C}.\par

Once EI is broadcasted to the network, the participating nodes in the blockchain start verifying the energy trade. First, the participants must locate the associated LP and EN. Recall that EI contains the  \textit{LP\_ID} and LP contains \textit{EN\_Ref} that is the identifier of the EN transaction. The verifiers first match the signatures and PKs in the transactions. The next step is for the verifiers to validate that the amount and price agreed in the transactions match. Once all the above steps are successfully validated,  EI and LP transactions are stored in the blockchain, which triggers the payment of the energy price to the energy producer. If the price in these transactions differs, the verifiers discard those transactions. \par

In case of detecting inconsistency in the amount of the injected energy in EI, the verifiers call a \textit{Dispute Resolution} (DR) smart contract. The DR smart contract contains rules to manage the situation where the energy producer failed to transfer the promised amount of energy to the consumer, for example, the energy produced by the solar panel of an energy producer may be less than the estimated production, which potentially impacts the traded energy. Based on the amount of transferred energy identified in EI, DR calculates a new energy price and generates a \textit{Price Update (PU)} transaction requesting the energy consumer to generate a new LP with exactly the same condition as the previous one while updating the price. PU is structured as  $<LP\_ID, Old\_Price, New\_Price>$. The new LP is broadcast to the network and is stored in the blockchain with the corresponding EI. \par

Recall that in the proposed framework, we defined the reputation factor that impacts the decision making of the nodes. The reputation is given by the DR smart contract based on the commitments of an energy producer. In the case of the above example, the DR will reduce the reputation of the node and inform all participants. In this study, we consider the negative reputation only, which is when a node misbehaved in the network.

\section{Case Studies}\label{sec: case study}
\begin{figure}
\centering
\tikzstyle{main} = [rectangle, minimum width=0.03cm, minimum height=1cm,text centered, draw=black,fill=black]

\tikzstyle{pro} = [rectangle, minimum width=0.15cm, minimum height=0.1cm, text centered, draw=red!80, fill=red!80]
\tikzstyle{con} = [circle, minimum width=0.1cm, text centered, draw=blue!50, fill=blue!80]
\tikzstyle{circ} = [circle, minimum width=0.5cm, text centered, draw=black]
\tikzstyle{arrow} = [thick,-,>=stealth]
\tikzstyle{int} = [circle, minimum width=0.1cm, text centered, draw=black, fill=black]

\tikzstyle{Mytext} = [rectangle, minimum width=0.5cm, minimum height=0.2cm,text centered]

\begin{tikzpicture}[node distance= 0.56 cm, auto] 
\footnotesize{
\centering
\node (main1) [main]{} ;

\node (circ1) [circ,right of=main1,xshift=0.3cm] {};
\node (circ2) [circ,right of=main1,xshift=0.1cm] {};
\node (int) [int, right of=main1,xshift=0.9cm]{} ;
\node (pro1) [pro,right of=int]{} ;
\node (pro2) [pro,right of=pro1] {};
\node (pro3) [pro,right of=pro2] {};
\node (con4) [con,right of=pro3]{} ;
\node (pro5) [pro,right of=con4] {};
\node (con6) [con,below of=pro5] {};
\node (pro7) [pro,below of=con6] {};
\node (pro8) [pro,below of=pro7]{} ;
\node (con9) [con,below of=pro8] {};
\node (con10) [con,right of=con9] {};
\node (pro11) [pro,right of=con10] {};
\node (con12) [con,right of=pro11] {};
\node (pro13) [pro,right of=con12]{} ;
\node (con14) [con,right of=pro13] {};
\node (pro15) [pro,right of=con14] {};
\node (pro16) [pro,right of=pro15] {};
\node (pro17) [pro,right of=pro16] {};
\node (con18) [con,below of=pro1] {};
\node (pro19) [pro,below of=con18]{} ;
\node (pro20) [pro,right of=pro19] {};
\node (con21) [con,right of=pro20] {};
\node (con22) [con,above of=pro2] {};
\node (pro23) [pro,above of=con22] {};
\node (con24) [con,right of=pro23] {};
\node (pro25) [pro,above of=pro5] {};
\node (con26) [con,above of=pro25] {};
\node (con27) [con,right of=con26] {};
\node (pro28) [pro,right of=con27] {};
\node (con29) [con,right of=pro28] {};
\node (pro30) [pro,right of=con29] {};
\node (con31) [con,right of=pro30] {};
\node (pro32) [pro,right of=con31] {};
\node (pro33) [pro,right of=pro5,xshift=1.5cm] {};
\node (text33) [Mytext,right of=pro33, xshift=0.2cm] {Consumer};
\node (con34) [con,below of=pro33,yshift=-0.3cm] {};
\node (text34) [Mytext,right of=con34, xshift=0.2cm] {Producer};

\node (text1) [Mytext,above of=pro1,yshift=-0.3cm] {1};
\node (text2) [Mytext,below of=pro2,yshift=0.3cm] {2};
\node (text3) [Mytext,below of=pro3,yshift=0.3cm] {3};
\node (text4) [Mytext,below of=con4,yshift=0.3cm] {4};
\node (text5) [Mytext,right of=pro5,xshift=-0.3cm] {5};
\node (text6) [Mytext,right of=con6,xshift=-0.3cm] {6};
\node (text7) [Mytext,right of=pro7,xshift=-0.3cm] {7};
\node (text8) [Mytext,right of=pro8,xshift=-0.3cm] {8};
\node (text9) [Mytext,below of=con9,yshift=0.3cm] {9};
\node (text10) [Mytext,below of=con10,yshift=0.3cm] {10};
\node (text11) [Mytext,below of=pro11,yshift=0.3cm] {11};
\node (text12) [Mytext,below of=con12,yshift=0.3cm] {12};
\node (text13) [Mytext,below of=pro13,yshift=0.3cm] {13};
\node (text14) [Mytext,below of=con14,yshift=0.3cm] {14};
\node (text15) [Mytext,below of=pro15,yshift=0.3cm] {15};
\node (text16) [Mytext,below of=pro16,yshift=0.3cm] {16};
\node (text17) [Mytext,below of=pro17,yshift=0.3cm] {17};
\node (text18) [Mytext,left of=con18,xshift=+0.25cm] {18};
\node (text19) [Mytext,left of=pro19,xshift=+0.25cm] {19};
\node (text20) [Mytext,below of=pro20,yshift=+0.3cm] {20};
\node (text21) [Mytext,below of=con21,yshift=+0.3cm] {21};
\node (text22) [Mytext,left of=con22,xshift=+0.25cm] {22};
\node (text23) [Mytext,left of=pro23,xshift=+0.25cm] {23};
\node (text24) [Mytext,below of=con24,yshift=+0.3cm] {24};
\node (text25) [Mytext,left of=pro25,xshift=+0.25cm] {25};
\node (text26) [Mytext,left of=con26,xshift=+0.25cm] {26};
\node (text27) [Mytext,below of=con27,yshift=+0.3cm] {27};
\node (text28) [Mytext,below of=pro28,yshift=+0.3cm] {28};
\node (text29) [Mytext,below of=con29,yshift=+0.3cm] {29};
\node (text30) [Mytext,below of=pro30,yshift=+0.3cm] {30};
\node (text31) [Mytext,below of=con31,yshift=+0.3cm] {31};
\node (text32) [Mytext,below of=pro32,yshift=+0.3cm] {32};

\draw [arrow] (main1) -- (circ2);
\draw [arrow] (circ1) -- (int);
\draw [arrow] (int) -- (pro1);
\draw [arrow] (pro1) -- (pro2);
\draw [arrow] (pro2) -- (pro3);
\draw [arrow] (pro3) -- (con4);
\draw [arrow] (con4) -- (pro5);
\draw [arrow] (pro5) -- (con6);
\draw [arrow] (con6) -- (pro7);
\draw [arrow] (pro7) -- (pro8);
\draw [arrow] (pro8) -- (con9);
\draw [arrow] (con9) -- (con10);
\draw [arrow] (con10) -- (pro11);
\draw [arrow] (pro11) -- (con12);
\draw [arrow] (con12) -- (pro13);
\draw [arrow] (pro13) -- (con14);
\draw [arrow] (con14) -- (pro15);
\draw [arrow] (pro15) -- (pro16);
\draw [arrow] (pro16) -- (pro17);
\draw [arrow] (pro2) -- (con22);
\draw [arrow] (con22) -- (pro23);
\draw [arrow] (pro23) -- (con24);
\draw [arrow] (pro5) -- (pro25);
\draw [arrow] (pro25) -- (con26);
\draw [arrow] (con26) -- (con27);
\draw [arrow] (con27) -- (pro28);
\draw [arrow] (pro28) -- (con29);
\draw [arrow] (con29) -- (pro30);
\draw [arrow] (pro30) -- (con31);
\draw [arrow] (con31) -- (pro32);
\draw [arrow] (pro1) -- (con18);
\draw [arrow] (con18) -- (pro19);
\draw [arrow] (pro19) -- (pro20);
\draw [arrow] (pro20) -- (con21);
}
\end{tikzpicture}
\caption{33-Bus test system.}\label{fig:test system}

\end{figure}

\begin{table}[]
\centering
\footnotesize{

\caption{Parameter Setup.}\label{tab:parameters}
\begin{tabular}{cccc}
\hline
\multicolumn{4}{c}{\textbf{Key parameters}} \\ \hline
Parameter & \multicolumn{1}{c|}{Value} & Parameter & Value \\ \hline
$P$ & \multicolumn{1}{c|}{14} & $C$ & 18 \\
$\rho_\lambda$ & \multicolumn{1}{c|}{0.01} & $\rho_\mu$ & 0.001 \\
 $\overline{\lambda}^G$& \multicolumn{1}{c|}{5 \cent/kWh^2} &$\underline{\lambda}^G$  & 25 \cent/kWh^2 \\
 $N$& \multicolumn{1}{c|}{2} & $\omega$ & 2 \cent/kWh/ km \\ \hline
\multicolumn{2}{c|}{\textbf{Producers' parameters}} & \multicolumn{2}{c}{\textbf{Consumers' parameters}} \\ \hline
$a_i$ & \multicolumn{1}{c|}{(0.5-1] \cent/kWh^2} & $a_j$ & (0.5-10] \cent/kWh^2 \\
 $b_i$& \multicolumn{1}{c|}{[5-10] \cent/kWh} & $b_j$ & [10-20] \cent/kWh \\
 $\underline{e}_i$& \multicolumn{1}{c|}{[0-5] kWh} & $\underline{e}_j$ &  [1-4] kWh\\
 $\overline{e}_i$& \multicolumn{1}{c|}{[5-10] kWh} & $\overline{e}_j$ & [6-10] kWh \\
 $\eta_i,\alpha_i,\beta_i$& \multicolumn{1}{c|}{[0-1]} & $\eta_j,\alpha_j,\beta_j$ & [0-1] \\ \hline
\end{tabular}
}
\end{table}

In this section, simulation case studies are provided to verify the operation of the proposed framework. As shown in Fig. \ref{fig:test system}, the considered test system is the IEEE 33-bus distribution system with 16 producers and 16 consumers. Table \ref{tab:parameters} summarizes key parameters and range of values for producers and consumers parameters.

\begin{figure}[tb!]
    \centering
    \includegraphics[scale=0.75]{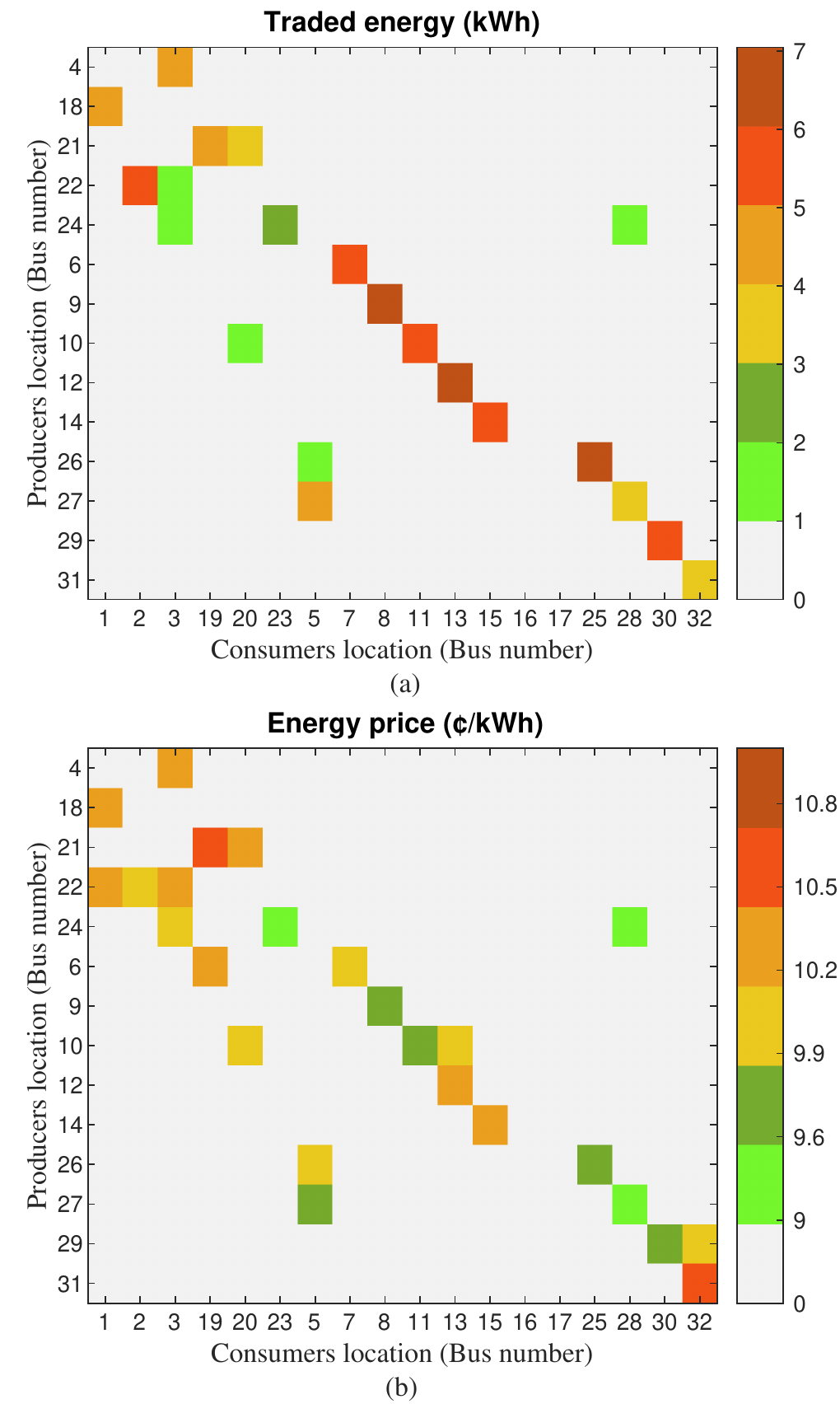}
    \caption{Transactions between producers and consumers; a) traded energy, b) energy price.}\label{fig: power and price res}
\end{figure}

Fig. \ref{fig: power and price res} illustrates the results of P2P trading in the test system. The traded energy and price in different transactions have various values based on the agents' preferences. Agents tend to trade energy with their closest neighbor agents to pay less grid service charge. For example, consumer at bus 1 buys energy from producer at bus  18. However, if the offer/ask from the agent at the nearest node is not available, or does not satisfy the requirement of the agents, they have to trade with other neighboring agents. For instance, while the nearest agents to agent 5 are agents at bus 4 and 6, this agent trades energy with producers at bus 26 and 27. Since agents 4 and 6 have preferred to trade with other neighboring nodes (agents at bus 3 and 7 respectively), their offers are not available to the agent 5. It can be seen that agents 16 and 17 do not buy any energy in the market. These agents have lower utility function parameters compared to their neighboring agents, which means that their willingness to pay for energy is less than agent 15, and hence, producer at bus 14 prefers to trade with the agent at bus 15.

\begin{figure}
    \centering
    \includegraphics[scale=0.85]{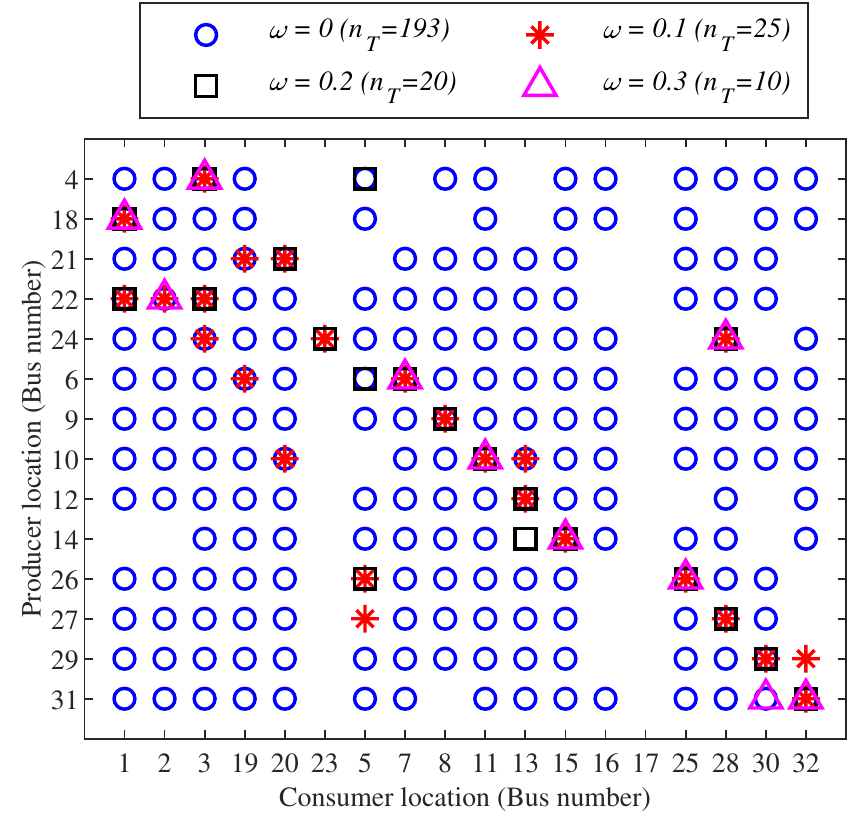}
    \caption{Impact of considering grid service charge on the number of transactions.}\label{fig:line flow}
   
\end{figure}

To investigate the impact of considering the grid service charge on the number of transactions ($n_T$), we implemented the energy trading algorithm for different values of $\omega$. The results are reported in Fig. \ref{fig:line flow}, where each marker indicates a transaction between the producer and consumer, and the total number of transactions in each case is given in parentheses. The case with \(\omega=0\) means that there is no limit on the distance of agents, and they can trade with any number of agents. Therefore, the number of transactions in this case is significantly higher. Increasing the value of $\omega$ reduces the number of transactions. The welfare of agents depend on the grid service charge that they pay (see (\ref{pro welf}) and (\ref{con welf})), and hence, increase in $\omega$ reduces their welfare as they have to trade less energy and pay more cost for utilizing the grid.

The negotiation among agents is an iterative process and the time required for agents to reach agreement depends on several factors including, number of agents, number of iterations required for convergence in Algorithm \ref{producer lag} and \ref{consumer alg}, the computation time required to solve (\ref{sell update dec}) and (\ref{buyer update dec}) by agents in each iteration, and communication time. The results of implementing the market settlement algorithm with and without implementing \textit{prioritization} step are compared, as reported in Table \ref{tab:prioritization}. The \textit{prioritization} step reduces the number of negotiating agents, and hence, reduces the number of communication in each iteration. On the other hand, agents need less time to solve (\ref{sell update dec}) and (\ref{buyer update dec}), as they have fewer decision variables after \textit{prioritization}. Therefore, applying \textit{prioritization} reduces negotiation time among agents by nearly 45\%.

\begin{table}[tb!]
\centering
\footnotesize{
\caption{Impact of Prioritization.}\label{tab:prioritization}
\setlength{\tabcolsep}{5pt}
\begin{tabular}{lcc}
\cline{2-3} \\[-0.7em]
 & \textbf{w prioritization} & \textbf{w/o prioritization} \\ \\[-0.7em] \hline
No. of decision variables & \multirow{2}{*}{38, 16} & \multirow{2}{*}{20, 6} \\ \\[-1em]
(producer, consumer) &  &  \\ \\[-1em]
No. of communications & \multirow{2}{*}{63} & \multirow{2}{*}{252} \\ \\[-1em]
(in each iteration) &  &  \\ \\[-1em]
No. of iterations for convergence & 163 & 206 \\ \\[-1em] \hline
Negotiation time (s) & 1.04 & 1.88 \\ \\[-1em] \hline
\end{tabular}
}
\end{table}

\begin{table}[tb!]
\centering
\footnotesize{

\caption{Comparative results of P2P market.}
\label{tab:p2p results}
\setlength{\tabcolsep}{5pt}
\begin{tabular}{lcc}
\cline{2-3} \\[-0.7em]
 & \textbf{P2P} & \textbf{No P2P} \\  \\[-0.7em] \hline
Total imported energy from grid (kWh) [$\sum_j{e_j^G}$] & 22.31 & 119 \\ \\[-1em]
Total exported energy to grid (kWh) [$\sum_i{e_i^G}$] & 8.46 & 105 \\ \\[-1em]
Total welfare of consumers ($\cent$) [$\sum_j{W_j}$] & 62.73 & -4143.04 \\ \\[-1em]
Total welfare of producers ($\cent$) [$\sum_i{W_i}$] & 242.64 & -302.03 \\ \\[-1em]
Total paid grid service charge ($\cent$) [$\sum_j \sum_i {e_{ij}^p \gamma_{ij}}$] & 50.44 & 0 \\ \hline
\end{tabular}
}
\end{table}
In order to demonstrate the efficacy of the P2P market, the results of this market are compared with the case that producers and consumers only trade with the grid. Comparative results are reported in Table \ref{tab:p2p results}. As it can be inferred from the results, P2P market reduces the imported and exported energy by agents to the grid, meaning they trade more with other P2P agents. Also, since the P2P market price is more beneficial for agents (see (\ref{price lim})), they can reach a higher value of welfare in the P2P market, though they have to pay a grid service charge to the grid operator.

As stated in Section \ref{sub:sec:advertisement}, in the proposed framework, the ATs are stored off  chain in AD. Here, we study the impact of using AD by evaluating the blockchain size and number of consensus rounds, i.e., evaluating how many times the validators shall run the consensus algorithm. Blockchain size shows the amount of saved storage space by employing AD while the number of consensus rounds evaluates the amount of computational overhead saved by running less consensus rounds. We employed the structure and configuration of the IEEE 33-bus distribution to implement a distributed network using Java programming language on Raspberry Pi 2. It is assumed that the size of each block is 10 transactions. The process involved in the consensus algorithm is abstracted out as it does not impact the functionality of the proposed method. Ten market intervals are implemented during which each energy producer generates an AT. To benchmark the results, a baseline method is considered where all ATs are stored in the blockchain. To focus on the impact of AD, we disregard the rest of the functions and assume ATs are the only transactions generated during each market interval. Based on the implementation results, the size of each AT is 1776 B. After 10 epochs, the baseline blockchain includes 16 blocks with the cumulative size of 314KB. Thus, each node must allocate 314KB storage space to store blockchain. Our solution off-loads this overhead to a central trusted node who is managing the AD, thus there is no memory overhead on the participating nodes in the blockchain. Assume $\nu $ represents the overhead associated with appending a new block that includes computational, packet and memory overhead. The proposed framework incurs no overhead during the \textit{advertisement} process on the validators while the overhead is   16$\nu$ in the conventional methods. We next evaluate the processing overhead associated with CoL. 

We proposed a CoL that enables the users to anonymously verify the location of the parties involved in energy trading. CoL enhances the anonymity level of users and thus protects user privacy. On the flip side, it increases the processing overhead on the users to generate and verify CoL. To evaluate the incurred overhead, we implemented our framework using Java programming language on Raspberry Pi 2, which represents low-resource devices. We measured the processing time for generating the CoL request, which involves generating a set of keys and forming the Merkle tree, and verifying the CoL, which involves verifying the existence of the PK in the Merkle tree and validating the corresponding signatures. The implementation results are shown in Table \ref{tab:COL-performance}. The verification of the CoL involves verifying two signatures, which potentially takes longer time than generating CoL. In addition to the processing overhead, CoL increases the size of the transactions. Table \ref{tab:COL-packet} compares the size of transactions and shows that CoL only affects the AT. It nearly doubles the size of AT, but this does not affect the size of the blockchain as AT's are stored off-chain only. All other transactions are unaffected by CoL. \par

\section{Security and Privacy Analysis}\label{sec:security}
In this section, we analyze the security and privacy of the proposed framework. We first outline threat mode and then discuss possible attacks and how to protect against those.\par 

\textbf{\textit{Threat Model:}} We assume that the adversary (or cooperative adversaries) can sniff the communications, discard transactions, generate fake transactions, and pretend to be another node in the network. The adversary may attempt to deanonymize a user by classifying blockchain transactions and monitoring real-time communications in blockchain. We assume standard secure encryption algorithms are in place, which cannot be compromised by the adversary. We assume smart meters are tamper resistance, and thus the end users cannot modify the transactions generated by the meters. \par 

\begin{table}[tb!]
\centering
\footnotesize{

\setlength{\tabcolsep}{5pt}
\caption{CoL processing time.}\label{tab:COL-performance}
\begin{tabular}{ccc}
\hline
 & CoL  formation  & CoL verification \\\hline
Processing time (ms) & 663.2 & 1795 \\\hline
\end{tabular}
}
\end{table}

\begin{table}[tb!]
\centering
\footnotesize{

\setlength{\tabcolsep}{5pt}
\caption{Comparison of transaction sizes.}\label{tab:COL-packet}
\begin{tabular}{ccccc}
\hline
 & AT  & EN & LP & EI \\\hline
Including  CoL (Bytes)  & 2193 & 1928 & 1056 & 1912 \\\hline
Excluding CoL (Bytes) & 1041  &  1928 & 1056 & 1912 \\\hline
\end{tabular}
}
\end{table}
\subsection{Security}
In the following, we discuss possible attacks and how the proposed framework protects against those. \par 

\textit{CoL Reply Attack:} In this attack, the malicious node attempts to employ CoL of another node to generate transactions.  The node that employs a CoL is required to sign the corresponding transaction with the private key corresponding to a PK that exists in MTR that is only known to the CoL generator. Thus, it is impossible for a malicious node to utilize the CoL of another node. \par 

\textit{Fake CoL:} In this attack, a malicious node pretends to be a genuine smart meter generates fake CoL that can later be used in its energy tradings. The CoL must be signed only by a genuine smart meter, and the CA validates the  PK of the verifier. In the case of this attack, CA will not validate PK, and thus the attack can be detected. \par 

\textit{Double selling:} In this attack, a malicious energy producer attempts to sell the same amount of energy to different consumers. Recall from Section \ref{sec:energy trading} that an energy trade involves three main transactions, which are EN, LP, and EI.  Once the agreed energy is injected to the grid, the smart meter of the energy producer generates a EI transaction that triggers the payment of the energy price to the producer.    The smart meter generates only one EI that includes a reference to the corresponding  LP, and LP includes a reference to the corresponding  EN. Thus, it is impossible for the energy producer to sell the same energy to multiple nodes. \par 
An energy producer may attempt to inject less energy than the agreed amount and claim the full price. The smart meter of the producer will only generate the EI if the full amount of agreed energy is injected to the grid. If the energy producer injects part of the energy and the expiry time approaches, the smart meter will generate an EI reflecting the amount that is injected to the grid. In this case, DR smart contract is called that adjusts the price of the energy and ensues the producer is only paid for the amount of energy injected to the grid. \par

\textit{Reputation Modification:} In this attack, the malicious node attempts to improve their reputation or reduce the reputation of another node in the network. Recall that blockchain is an immutable ledger that makes it impossible to modify or remove previously stored transactions, which makes it impossible for the attacker to modify their reputation. To reduce the reputation of another node, the malicious node shall modify the code of the smart contract, which is impossible due to the immutability of the blockchain. DR smart contract is the only entity that can reduce the reputation of a node. All participants know the address of the valid DR contract. When participating nodes receive reputation reduction from a contract, they first verify if the contract address matches with the genuine DR smart contract. If so, they accept the new reputation. Otherwise, they discard the transaction. 

\subsection{Privacy}
In the following, we analyze the proposed framework from the privacy perspective. Recall from Section \ref{sec:energy trading} that the grid operator charges a grid service charge per each transaction that depends on the distance between the energy consumer and producer. Thus, the consumer and producer must prove their location, however, this may compromise their privacy as malicious nodes can classify the blockchain transactions to deanonymize the user. To address this challenge, we proposed A-PoL that enables the participants in the blockchain to verify the location of an anonymous smart meter using a CoL. Assume node \textit{A} is using A-PoL. The privacy of \textit{A} can be studied from the perspective of the following entities: i) CA: \textit{A} uses the PK populated by the CA only to prove its identity to the verifier. CoL employed by \textit{A} includes PK of the verifier and not \textit{A}. Given that the verifier is selected randomly by \textit{A} and there is no link between \textit{A} and the verifier, the CA is unable to identify the transactions generated by \textit{A}, ii) verifier: \textit{A} only sends MTR to the verifier that hides the actual PKs of \textit{A} from the verifier. \textit{A} reveals the PKs in the Merkle tree to prove ownership of CoL. A group of smart meters may choose to create a single MTR, which further protects their privacy, and iii) network participants: the network participants only receive CoL that contains of the verifier and MTR. As outlined earlier, there is no link between the verifier and \textit{A}, thus knowledge of the identity of the verifier does not impact the privacy of \textit{A}. The Merkle tree includes a number of PKs that are employed by \textit{A} (or other involved smart meters) to generate transactions, thus, \textit{A} may generate multiple transactions with the same PK. This potentially reduces the anonymity level of the user as malicious nodes may attempt to deanonymize a user by classifying their transactions. The anonymity level of \textit{A} largely depends on the number of PKs employed in the Merkle tree. The large number of PKs incur overheads on \textit{A} to manage the keys. Thus, there is a trade-off between the number of keys in the Merkle tree and user anonymity. \par 

Recall from Section \ref{sub:sec:market-settlement} that the energy producer and consumer employ a cost/utility function, as shown in  (\ref{cost-func-producer}) and (\ref{cost-function-consumer}), which represent their willingness to pay or accept energy price based on their preferences and concerns. These functions depend on $a_i$, $b_i$, and $c_i$, and thus it is critical for the producers and consumers to keep these values private. In the proposed framework, the market settlement does not need the nodes to reveal $a_i$, $b_i$ and $c_i$, which in turn enhances the privacy of the users. 

\section{Conclusion and Future Works}\label{sec:conclusion}
In this paper, we propose a blockchain-enabled P2P energy market, which provides a secure and privacy-preserving environment for energy trading between producers and consumers. A decentralized market settlement process is designed, which allows agents to trade energy without revealing their private information. The grid service charge, calculated based on the distance between producer and consumer, is used to incite agents trade energy locally and to reduce the possibility of overloading electricity grid lines.\par
To reduce the blockchain memory footprint, we propose AD that stores the energy advertisements and is maintained by the grid operator. A \textit{prioritization} step is implemented to enable agents to select their trading partners based on their location and reputation factors. In order to allow agents to prove their location without revealing their real identity, an A-PoL algorithm is proposed using CoL issued by smart meters. Simulation results on IEEE 33-bus test system confirm that the proposed framework improves the welfare of agents through P2P trading, while their privacy is protected. Furthermore, employing AD to store ATs, and limiting the number of trading partners through \textit{prioritization} decrease the system overheads.   

For future work is needed to relax the tamper resistance assumption considered for smart meters. Relaxing this assumption complicates the trust issue as the smart meters may generate fake transactions. As another research direction, the impact of mobility of the smart meters on A-PoL can be studied. In such cases, the CA must ensure that the location of a meter is as claimed before granting a certificate. It is critical for the CA to be able to revoke granted certificates as  smart meters may change its location. Another challenge for future work is to explore ways of decentralizing the AD without increasing the blockchain memory footprint to achieve an even more decentralized energy marketplace.

\bibliographystyle{IEEEtran}
 \bibliography{ref}

\end{document}